\documentclass{aa}
\usepackage{natbib}
\bibpunct{(}{)}{;}{a}{}{,}
\usepackage{graphicx}

\begin{document}

\title{Stationary equatorial MHD flows in general relativity}

\author{F. Daigne\thanks{\emph{Present address:} Service
    d'Astrophysique, CEA/Saclay, 91191 Gif sur Yvette Cedex, France}
  \and G. Drenkhahn}

\institute{Max-Planck-Institut f\"ur Astrophysik, Postfach~1317,
  85741~Garching bei M\"unchen, Germany}

\mail{daigne@discovery.saclay.cea.fr}

\date{Received ... / Accepted ...} 

\abstract{We derive a new formulation of the fully general
  relativistic equations describing a stationary equatorial MHD
  outflow from a rotating central object.  The wind solution appears
  as a level contour of a ``Bernoulli'' function fixed by the
  requirements that it must pass through the slow and fast critical
  points.  This approach is the general relativistic extension to the
  classical treatment of \protect\citet{sakurai:85}.  We discuss in
  details how the efficiency of the magnetic to kinetic energy
  conversion depends mainly on the geometry of the flux tubes and show
  that the magnetic acceleration can work very well under some
  conditions. We show how this tool can be used for the study of
  several astrophysical phenomena, among which gamma--ray bursts.
  }

\maketitle

\section{Introduction}

Magnetized winds are believed to be present in many astrophysical
objects. They were first put forward in the context of the solar wind.
More recently they were discovered to probably play a major role in
many situations where a relativistic flow is powered by a central
rapidly rotating compact object: wind of pulsars, jets in radio
galaxies, quasars, Seyfert galaxies and BL Lac objects, microquasars
and even possibly gamma--ray bursts.

The first quantitative model of a magnetic stellar wind was developed
by \citet{weber:67}. They were considering the equations of a
stationary, axisymmetric, polytropic flow near the equatorial plane in
classical MHD. They found that such a wind can carry off most of the
angular momentum of the star and are very efficient to accelerate
particles up to very high velocities. An important feature of magnetic
winds is the existence of three ``critical points'' where the velocity
of the flow equals the wave velocity of the three MHD wave modes (the
slow, Alfv\'en and fast modes) whereas in comparison non-magnetic
winds \citep{parker:58} have only one ``critical point'' where the
velocity of the flow equals the sound speed (sonic waves being the
only present wave mode).

The first extension of this theory to relativistic winds is due to
\citet{michel:69} in the context of radio pulsars. This work was
considering cold outflows driven by rapidly rotating highly magnetized
neutron stars. The main conclusion was that the efficiency of the
magnetic to kinetic energy conversion was extremely low compared to
the classical case. \citet{goldreich:70} studied cool isothermal
relativistic winds. \citet{kennel:83} extended Michel's model to
finite temperatures and relativistic injection speeds. All these works
were always limited to the equatorial plane and either completely
neglected the effect of gravity or adopted an approximative treatment
for it. \citet{okamoto:78} first included an exact general
relativistic description of the gravity field. His work was also not
limited to the equatorial plane, applying for that the powerful
concept of flux tubes. However he restricted his study to pressureless
flows only. In a series of papers
\citet{camenzind:86a,camenzind:86b,camenzind:87} derived a complete
set of equations describing a stationary axisymmetric relativistic
magnetic wind in an arbitrary metric. He then solved these equations
in some particular cases (cold flows, jet geometries).

The goal of this paper is to present a formulation of the equations
governing a stationary axisymmetric MHD flow in the equatorial plane
including an exact treatment of all effects (thermal pressure, gravity
and arbitrary shapes of flux tubes) which allows a direct comparison
with the classical model of \citet{weber:67}, so that the relativistic
effects can be easily identified.  This is done in
Sects.~\ref{sec:windeq} and \ref{sec:descr}, where we worked by
analogy with the formulation of the classical case by
\citet{sakurai:85}. Then we study in details the efficiency of the
magnetic to energy conversion (Sect.~\ref{sec:eff}), in particular the
influence of the flux tubes geometry and of the gravity. We confirm
and extend the results of \citet{begelman:94} and we show that a large
variety of situations is expected from very inefficient winds like
those considered by \citet{michel:69} to highly efficient cases.
Because our model assumes axisymmetry and focus on the equatorial
plane, it fully applies only to simple astrophysical objects like
isolated neutron stars.  On the other hand it can also describe the
outer parts of more complex objects, e.g. compact objects with
accretion disks or complex magnetospheres, as long as the magnetic
field can be approximated as monopole like at these distances from the
source.  In the particular case of gamma--ray bursts the possibility
of Poynting-flux dominated fireballs is briefly discussed in
Sect.~\ref{sec:grb}.  This is summarized in
Sect.~\ref{sec:conclusions}.

\section{The wind equations}
\label{sec:windeq}

The conservation laws of the general-relativistic magnetohydrodynamics
have been derived by \citet{bekenstein:78}.  \citet{camenzind:86b}
used these results to obtain the equations governing a stationary
axisymmetric wind.  In this section, we first recall these equations
and we apply them to the particular case of a flow occurring in the
equatorial plane.  Then we derive a new formulation for this problem
where the wind solution is a level contour of a Bernoulli-like
function.  A very similar formulation was earlier studied by
\citet{sakurai:85} for the non-relativistic case.  The similarity
allows us to compare our results with those of the classical case.

\subsection{Assumptions and basic equations}

Any stationary and axisymmetric space-time can be represented by the
following metric:
\begin{equation}
  \mathrm{d}s^2 = g_{tt} \mathrm{d}t^2 
  + 2 g_{t\phi} \mathrm{d}t \mathrm{d}\phi
  + g_{\phi\phi} \mathrm{d}\phi^2
  + g_{ab} \mathrm{d}x^a \mathrm{d}x^b
\end{equation}
where the coordinates $t$ and $\phi$ correspond to the two symmetries
of the space-time defined by the two Killing fields
$k=\partial_\mathrm{t}$ (stationarity) and $m=\partial_\phi$
(axisymmetry). The metric coefficients $g_{ab}$ depend only on the two
remaining coordinates $x^a$ ($a=1,2$).  Note that in the whole paper
we will make use of the $(-+++)$ signature for the metric.  The
electromagnetic field is described by the field tensor $F^{\mu\nu}$
and the dual field tensor $F^{*\mu\nu} =
\frac{1}{2}\epsilon^{\mu\nu\rho\sigma} F_{\rho\sigma}$
($\epsilon^{\mu\nu\rho\sigma}$ being the Levi-Civita alternating
tensor) which satisfy the Maxwell equations
\begin{equation}
  \nabla_\mu F^{*\mu\nu} = 0\ .
\end{equation}
The motion of the plasma is governed by the energy- and momentum
conservation equation
\begin{equation}
  \nabla_\mu T^{\mu\nu} = 0\ ,
\end{equation}
where the energy-momentum tensor $T^{\mu\nu}$ is made up of the fluid
part
\begin{equation}
  T^{\mu\nu}_\mathrm{matter} 
  = \rho \frac{h}{c^2} u^\mu u^\nu + P g^{\mu\nu}
\end{equation}
 and of the electromagnetic part
\begin{equation}
  T^{\mu\nu}_\mathrm{em}
  = \frac{1}{4\pi}\left( 
    \frac{b^2}{c^2} u^\mu u^\nu 
    + \frac{b^2}{2} g^{\mu\nu} - b^\nu b^\mu 
  \right)\ .
\end{equation}
$b^\mu$ is the magnetic field according to a comoving observer,
written as
\begin{equation}
  b^\mu = F^{*\mu\nu} u_\nu
\end{equation}
with $b^2=b_\mu b^\mu$.  In the comoving frame $b^\mu$ reduces to the
common magnetic field
\begin{equation}
  \label{eq:Bco}
  b^\mu_\mathrm{co}
  = F^{*\mu0}
  = (0,\vec B)\ . 
\end{equation}
All dissipative effects (heat conduction, viscosity, cooling by
radiation, etc.) have been neglected so that the flow is adiabatic.
Ideal MHD is also assumed, which means that the proper electric field
as seen in the plasma frame vanishes
\begin{equation}
  F^{\mu\nu} u_\mu = 0\ .
\end{equation}
The following quantities appear in these equations: $u^\mu$ is the
4-velocity, $\rho$ is the comoving mass density, $P$ is the pressure
and $h$ is the specific enthalpy, which is given by (assuming a
constant adiabatic index $\gamma$)
\begin{equation}
  h = c^2 + \frac{\gamma}{\gamma-1}\frac{P}{\rho}\ .
\end{equation}
The last assumption is that the particle number (or mass) is conserved
\begin{equation}
  \label{eqMassConservation}
  \nabla_\mu \left(\rho u^\mu\right) = 0\ .
\end{equation}

Before writing the wind equations, it is useful to define the specific
angular momentum of the flow
\begin{equation}
  l = -\frac{u_\phi}{u_t} c^2\ .
\end{equation}
As a consequence of the symmetries, the non vanishing elements of the
electromagnetic tensor are given by
\begin{eqnarray}
  F_{t1} & = & - F_{1t}
  =  c \Omega(P)\frac{\rho}{\eta(P)}\sqrt{-g} u^2\ ,\\
  F_{t2} & = & - F_{2t} 
  = -c \Omega(P)\frac{\rho}{\eta(P)}\sqrt{-g} u^1\ ,\\
  F_{\phi1} & = & -F_{1\phi} 
  = -c \frac{\rho}{\eta(P)}\sqrt{-g} u^2\ ,\\
  F_{\phi2} & = & -F_{1\phi} 
  = -c \frac{\rho}{\eta(P)}\sqrt{-g} u^{1}\ ,\\
  F_{12} & = & F_{21} 
  = c \frac{\rho}{\eta(P)}\sqrt{-g} 
  \left(\Omega(P) u^t - u^\phi \right)\ ,
\end{eqnarray}
where the angular frequency of the streamline $\Omega(P)$ at its
footpoint and the mass flux per unit flux tube $\eta(P)$ are constant
along each flow line $P$.  The corresponding magnetic field is
\begin{eqnarray}
  b^t & = & \frac{\rho}{\eta(P)} \left[
    1+\frac{u_t u^t}{c^2} \left(
      1-\frac{\Omega(P)l}{c^2}
    \right)
  \right]\ ,\\
  b^\phi & = & \frac{\rho}{\eta(P)} \left[
    \Omega(P) + \frac{u_t u^\phi}{c^2} \left(
      1-\frac{\Omega(P) l}{c^2}
    \right)
  \right]\ ,\\
  b^1 & = & \frac{\rho}{\eta(P)} \frac{u_t}{c^2}
  \left(1-\frac{\Omega(P)l}{c^2}\right) u^1\ ,\\
  b^2 & = & \frac{\rho}{\eta(P)} \frac{u_t}{c^2}
  \left(1-\frac{\Omega(P)l}{c^2}\right) u^2\ .
\end{eqnarray}
It is also useful to give the expression of the classical magnetic
field in the frame of the central object for the comparison with the
classical case. In this frame, it is related to the dual field tensor
by $B_\mu=F_{*\mu0}$. Then
\begin{eqnarray}
  B^t & = &
  -\frac{1}{c^2}\frac{\rho}{\eta(P)}
  \left(\Omega(P)u^t-u^\phi\right) g_{t\phi}\ ,\\
  B^\phi & = & 
  \frac{1}{c^2}\frac{\rho}{\eta(P)}
  \left(\Omega(P)u^t-u^\phi\right) g_{tt}\ ,\\
  B^1 & = & 
  -\frac{1}{c^2}\frac{\rho}{\eta(P)}
  \left( g_{tt} + \Omega(P) g_{t\phi} \right) u^1\ ,\\
  B^2 & = & 
  -\frac{1}{c^2}\frac{\rho}{\eta(P)}
  \left( g_{tt} + \Omega(P) g_{t\phi} \right) u^2\ .
\end{eqnarray}

Because the flow is stationary and axisymmetric, the total angular
momentum $L(P)$ and the total energy $E_\mathrm{tot}(P)$ are also
conserved along each flow line $P$ which provides us with two new
equations
\begin{eqnarray}
  L(P) &=& -\frac{u_t}{c^2}\cdot
  \biggl\lbrace \frac{h}{c^2} l
  + \frac{\rho}{4\pi\eta^2(P)}
  \nonumber\\
  \label{eqMomentum}
  &&\cdot\left[
      \frac{ g_{tt} + \Omega(P) g_{t\phi}}{c^2} l
      +\left( g_{t\phi} + \Omega(P) g_{\phi\phi} \right)
    \right]
  \biggr\rbrace
\end{eqnarray}
and
\begin{eqnarray}
  E_\mathrm{tot}(P) 
  &=& -\frac{u_t}{c^2}\cdot\biggl\lbrace
  h + \Omega(P) \frac{\rho}{4\pi\eta^2(P)}
  \nonumber\\
  &&\cdot\left[
    \frac{ g_{tt} + \Omega(P) g_{t\phi}}{c^2} l
    + \left( g_{t\phi} + \Omega(P) g_{\phi\phi}\right)
  \right]\biggr\rbrace .
\end{eqnarray}
It is convenient to write the total energy as $E_\mathrm{tot}(P) =
c^2+E(P)+\Omega(P)L(P)$ so that the energy conservation can
have the simpler form:
\begin{equation}
  \label{eqEnergy}
  E(P) + c^2 =
  -h \frac{u_t}{c^2}
  \left[1-\frac{\Omega(P)l}{c^2}\right]\ .
\end{equation}
Each flow line $P$ is completely determined by the four constants
$\Omega(P)$, $\eta(P)$, $L(P)$ and $E(P)$. The light surface is
defined by
\begin{equation}
  \label{eqLC}
  g_{tt} + 2 g_{t\phi}\Omega(P) + g_{\phi\phi}\Omega^2(P) = 0
\end{equation}
and the Alfv\'en point is fixed by two conditions
\begin{eqnarray}
  \frac{1}{c^2} \left[
    g_{tt} + 2 g_{t\phi}\Omega(P) + g_{\phi\phi}\Omega^2(P)
  \right]_\mathrm{A} & = & 
  - M_\mathrm{A}^2\ ,\label{eqAP}\\
  \frac{1}{c^2}
  \frac{\left( g_{t\phi} + \Omega(P) g_{\phi\phi}\right)_\mathrm{A}}
  {M_\mathrm{A}^2} & = &
  \frac{L(P)}{E(P)+c^2}\ ,
\end{eqnarray}
where the ``Mach'' number $M$ is given by
\begin{equation}
  M^2 = \frac{4\pi\eta^2(P)\,h}{\rho c^2}\ .
\end{equation}
One sees immediately from (\ref{eqLC}) and (\ref{eqAP}) that the
Alfv\'en point stays always inside the light surface (because of
$M_\mathrm{A}>0$).

\subsection{The wind equations in the equatorial plane}

We use now the spherical coordinates ($x^1=r$ and $x^2=\theta$) and
limit our study to the equatorial plane $\theta=\frac{\pi}{2}$.  The
specification of the flow line $P$ is dropped from here on and
$\Omega$, $\eta$, $E$, and $L$ are used instead of $\Omega(P)$,
$\eta(P)$, $E(P)$, and $L(P)$.  Because of the symmetry, $u^\theta$
and $B^\theta$ vanish in this plane but this is not necessarily the
case for their derivatives. Then the conservation of mass
(\ref{eqMassConservation}) can be written
\begin{equation}
  \label{eqM}
  \sqrt{-g} s(r) \rho u^r = \dot{m}\ ,
\end{equation}
where the function $s(r)$ depends on the geometry of the flux tubes.
In the simple case where $\partial_\theta u^\theta = 0$ and
$\partial_\theta B^\theta=0$, we have $s(r)=\mathrm{const.}$ (constant
opening angle). Otherwise we have
\begin{equation}
  \left.
    \partial_\theta u^\theta
  \right|_{\theta=\frac{\pi}{2}}
  = \frac{s'(r)}{s(r)}u^r\ .
\end{equation}
The conservation of angular momentum (\ref{eqMomentum}) and the
conservation of energy (\ref{eqEnergy}) read
\begin{eqnarray}
  L &=&
  \left[
    g_{\phi\phi}\frac{h}{c^2} 
    - \frac{g_{t\phi}^2 - g_{tt} g_{\phi\phi}}{c^2}
    \frac{\Phi^2\rho}{4\pi\dot{m}^2}
  \right] u^\phi
  \nonumber\\
  && + \left[
    g_{t\phi} \frac{h}{c^2}
    + \Omega \frac{g_{t\phi}^2 - g_{tt} g_{\phi\phi}}{c^2}
    \frac{\Phi^2\rho}{4\pi\dot{m}^2}
  \right] u^t
  \label{eqL}
\end{eqnarray}
and
\begin{eqnarray}
  E + c^2 &=&
  -h \left[\frac{g_{tt} + \Omega g_{t\phi}}{c^2} u^t
    + \frac{g_{t\phi} + \Omega g_{\phi\phi}}{c^2} u^\phi
     \right]\ ,
  \label{eqE}
\end{eqnarray}
where instead of using $\eta$ we have introduced the magnetic flux
$\Phi=\dot{m}/\eta$. The Eqs.~(\ref{eqM}), (\ref{eqL}) and (\ref{eqE})
are completed by the normalization of the four velocity
\begin{equation}
  g_{tt} \left( u^t \right)^2
  + 2 g_{t\phi} u^t u^\phi
  + g_{\phi\phi} \left(u^\phi\right)^2
  + g_\mathrm{rr} \left(u^r\right)^2 
  = -c^2\label{eqV}
\end{equation}
and the equation of state. We assume here, like in \citet{sakurai:85},
a polytropic relation $P=\kappa \rho^\gamma$ so that the specific
enthalpy is given by
\begin{equation}
  \label{eqH}
  h = c^2 + \frac{\gamma}{\gamma-1}\kappa\rho^{\gamma-1}\ .
\end{equation}
The system of Eqs.~(\ref{eqM}), (\ref{eqL}), (\ref{eqE}), (\ref{eqV})
and (\ref{eqH}) describes entirely the flow determined by the six
constants $\Omega,E,L,\Phi,\dot{m},\kappa$ and the free function
$s(r)$. In addition the two supplementary conditions
\begin{eqnarray}
  \label{eqA1}
  -\frac{1}{c^2} \left(
    g_{tt} + 2\Omega g_{t\phi}
    + \Omega^2 g_{\phi\phi}
  \right)_\mathrm{A} 
  &=& M_\mathrm{A}^2\nonumber\\
  &=& 
  \frac{4\pi\dot{m}^2}{\Phi^2}
  \frac{h_\mathrm{A}}{\rho_\mathrm{A}c^2}
\end{eqnarray}
and
\begin{equation}
  \label{eqA2}
  - \frac{1}{c^2} \left(
    \frac{ g_{t\phi} + \Omega g_{\phi\phi}}
    { g_{tt} + 2\Omega g_{t\phi}
      + \Omega^2 g_{\phi\phi}}
  \right)_\mathrm{A} 
  = \frac{L}{E+c^2}  
\end{equation}
(all quantities with index A are computed at the Alfv\'en point) must
be fulfilled, so that the flow remains regular at the Alfv\'en point.

The classical limit (for a weak gravitational field and for velocities
small compared to the speed of light) of this system of equations in
the case where $s(r)=\mathrm{const.}=1$ gives exactly the Eqs.~(1) to
(6) in \citet{sakurai:85}. Notice that $E$, $\Phi$ and $\Omega$ have
the same meaning in both papers whereas we use here different
notations for the mass flux $\dot{m}$, the total angular momentum $L$
and the polytropic constant $\kappa$ which are respectively $f$,
$\Omega r_\mathrm{A}^2$ and $K$ in \citet{sakurai:85}.  Notice also
that the relation~(\ref{eqA2}) between $L$ and $r_\mathrm{A}$ tends
towards $L=\Omega r_\mathrm{A}^2$ in the classical limit, so that all
notations are fully consistent.

\subsection{A Bernoulli-like formulation}

Following \citet{sakurai:85} we use the dimensionless variables
$x=r/r_\mathrm{A}$ and $y=\rho/\rho_\mathrm{A}$. The metric
coefficients are also normalized to become dimensionless $\tilde
g_{tt}(x)= g_{tt}/c^2$, $\tilde g_{t\phi}(x)= g_{t\phi}/(c r)$,
$\tilde g_{\phi\phi}(x)= g_{\phi\phi}/r^2$, $\tilde{g}_{rr}(x)=g_{rr}$
and $\sqrt{-\tilde{g}}(x)=\sqrt{-g}/r^2$.  These coefficients may be
not only functions of $x$ but also of some parameters defining the
metric (they are given for the Minkowski, Schwarzschild and Kerr
metric in appendix~\ref{appMetric}). We define $\tilde{\varpi}^2 =
\left(\tilde g_{t\phi}^2-\tilde g_{tt}\tilde
  g_{\phi\phi}\right)/c^2$.  We normalize $s$ by
$\tilde{s}(x)=s/s_\mathrm{A}$ and define a dimensionless specific
enthalpy by $\tilde{h}=h/c^2$.  Concerning four-vectors like $u^\mu$
or $B^\mu$ we will use the definitions $\tilde{A}^t=A^t$,
$\tilde{A}^\phi=r\sin{\theta} A^\phi$, $\tilde{A}^r=A^r$ and
$\tilde{A}^\theta=r A^\theta$ so that the spatial part of these
vectors is now given in the usual basis $\partial_r,
\frac{1}{r}\partial_\theta, \frac{1}{r\sin{\theta}}\partial_\phi$,
where all components have the same dimension.  We introduce four
normalized parameters
\begin{eqnarray}
  \beta' &=&
  \frac{1}{c^2}\left(
    \frac{\dot{m}}{s_\mathrm{A}r_\mathrm{A}^2\rho_\mathrm{A}}
  \right)^2\ ,\\
  \Theta' &=& \frac{\gamma\kappa\rho_\mathrm{A}^{\gamma-1}}{c^2}\ ,\\
  \omega' &=& \frac{\left(\Omega r_\mathrm{A}\right)^2}{c^2}\ ,\\
  E' &=& \frac{E}{c^2}\ ,
\end{eqnarray}
and we are now able to rewrite the system of Eqs.~(\ref{eqM}) and
(\ref{eqL}) to (\ref{eqH}):
\begin{eqnarray}
  \sqrt{\beta'} &=& 
  \sqrt{-\tilde{g}}\ \tilde{s}\ x^2 y \frac{\tilde{u}^r}{c}
  \label{eqNM}\\
  \frac{K_\mathrm{A}}{M_\mathrm{A}^2}\left(E'+1\right) &=&
  x\left\lbrace
    \left[
      \tilde g_{\phi\phi}\tilde{h}
      - \frac{\tilde{h}(1)}{M_\mathrm{A}^2}\tilde{\varpi}^2 y
    \right] 
    \frac{\tilde{u}^\phi}{c}\right. 
  \nonumber\\
  &&\left.+\left[
      \tilde g_{t\phi}\tilde{h}
      + \sqrt{\omega'}\frac{\tilde{h}(1)}{M_\mathrm{A}^2}
      \tilde{\varpi}^2 x y\right]
    \tilde{u}^t
  \right\rbrace
  \label{eqNL}\\
  E'+1 &=&
  - \tilde{h}\left[\left(
      \tilde g_{tt} + \sqrt{\omega'}\tilde g_{t\phi} x
    \right)\tilde{u}^t\right.
  \nonumber\\
  &&\left.+\left(
      \tilde g_{t\phi} + \sqrt{\omega'}\tilde g_{\phi\phi} x
    \right)\frac{\tilde{u}^\phi}{c}\right]
  \label{eqNE}\\
  -1 &=&
  \tilde g_{tt} \left(\tilde{u}^t\right)^2
  + 2\tilde g_{t\phi} \tilde{u}^t\frac{\tilde{u}^\phi}{c}
  + \tilde g_{\phi\phi}\left(\frac{\tilde{u}^\phi}{c}\right)^2
  \nonumber\\&&
  + \tilde{g}_{rr}\left(\frac{\tilde{u}^r}{c}\right)^2
  \label{eqNV}\\
  \tilde{h}(y) &=&
  1+\frac{\Theta}{\gamma-1}y^{\gamma-1}
  \label{eqNH}
\end{eqnarray}
In (\ref{eqNE}) and (\ref{eqNL}) the constants $L/r_\mathrm{A}c$ and
$\Phi^2\rho_\mathrm{A}/4\pi\dot{m}^2$ have been eliminated using the
conditions (\ref{eqA1}) and (\ref{eqA2}) at the Alfv\'en point, which
now read
\begin{eqnarray}
  -\left(
    \tilde g_{tt}(1) + 2\sqrt{\omega'}\tilde g_{t\phi}(1) 
    + \omega'\tilde g_{\phi\phi}(1)
  \right)
  &=& M_\mathrm{A}^2\nonumber\\
  &=& \frac{4\pi\dot{m}^2}{\Phi^2}
  \frac{\tilde{h}(1)}{\rho_\mathrm{A}}\\  
\end{eqnarray}
and
\begin{equation}
  - \frac{\tilde g_{t\phi}(1) + \sqrt{\omega'}\tilde g_{\phi\phi}(1)}
  {M_\mathrm{A}^2}
  = \frac{L}{r_\mathrm{A} c}\frac{1}{E'+1}
\end{equation}
so that
\begin{equation}
  \frac{\Phi^2\rho_\mathrm{A}}{4\pi\dot{m}^2} 
  = \frac{\tilde{h}(1)}{M_\mathrm{A}^2}
  \qquad\mbox{and}\qquad
  \frac{L}{r_\mathrm{A} c} 
  = \frac{K_\mathrm{A}}{M_\mathrm{A}^2}\left(E'+1\right)  
\end{equation}
with $K_\mathrm{A} = \tilde g_{t\phi}(1) + \sqrt{\omega'}\tilde
g_{\phi\phi}(1)$.

The component $\tilde{u}^{r}$ can be expressed from (\ref{eqNM}) and
substituted into (\ref{eqNV}) to provide a first relation between
$\tilde{u}^t$ and $\tilde{u}^\phi$.  A second relation between these
two components is given by (\ref{eqNL}) when subtracting
$K_\mathrm{A}/M_\mathrm{A}^2\times$~(\ref{eqNE}).  It allows us to
express all components of the four velocity as functions of only $x$
and $y$. Then the remaining Eq.~(\ref{eqNE}) becomes a Bernoulli-like
equation $\tilde{H}(x,y)=\mathrm{const.}$ as (10) of
\citet{sakurai:85}.  We do not further elaborate on the different
steps which lead to this final expressions:
\begin{eqnarray}
  \tilde{h}(y) 
  &=& 1+\frac{\Theta'}{\gamma-1}y^{\gamma-1}\ ,\\
  \tilde{u}^t  
  &=& \sqrt{K(x,y)}\frac{N(x,y)}{\sqrt{\mathcal{D}(x,y)}}\ ,\\
  \frac{\tilde{u}^\phi}{c} 
  &=& \sqrt{K(x,y)}\frac{D(x,y)}{\sqrt{\mathcal{D}(x,y)}}\ ,\\
  \frac{\tilde{u}^r}{c} 
  &=& \frac{\sqrt{\beta'}}{\sqrt{-\tilde{g}}\tilde{s}x^2y}\ ,\\
  \tilde{H}(x,y)+1
  &=& \tilde{\varpi}^2 x \tilde{h}(y)\,\sqrt{K(x,y)}
  \frac{\left|\mathcal{N}(x,y)\right|}{\sqrt{\mathcal{D}(x,y)}}
  \nonumber\\
  &=& E'+1\ ,
  \label{eqBernoulli}
\end{eqnarray}
where we have introduced the following auxiliary functions:
\begin{eqnarray}
  K(x,y) 
  &=& 1 + \frac{\tilde{g}_{rr}}{-\tilde{g}}
  \frac{\beta'}{\tilde{s}^2 x^4 y^2}\ ,\\
  N(x,y)
  &=& \left[
    \left(M_\mathrm{A}^2+\sqrt{\omega'}K_\mathrm{A}\right)
    \tilde g_{\phi\phi}\ x 
    + K_\mathrm{A} \tilde g_{t\phi} 
  \right] \tilde{h}(y)
  \nonumber\\
  & & -\tilde{h}(1)\tilde{\varpi}^2 x y\ ,\\
  D(x,y) 
  &=& -\left[ 
    \left(M_\mathrm{A}^2+\sqrt{\omega'}K_\mathrm{A}\right)
    \tilde g_{t\phi}\ x 
    + K_\mathrm{A} \tilde g_{tt}
  \right] \tilde{h}(y)
  \nonumber\\
  & & -\sqrt{\omega'}\tilde{h}(1)\tilde{\varpi}^2 x^2 y\ ,
  \nonumber\\
  \mathcal{N}(x,y)
  &=& \left(
    \tilde g_{tt} + 2\sqrt{\omega'}x\ \tilde g_{t\phi} 
    + \omega' x^2\ \tilde g_{\phi\phi}
  \right) \tilde{h}(1) y
  \nonumber\\
  & & + M_\mathrm{A}^2\tilde{h}(y)\ ,\\
  \mathcal{D}(x,y)
  &=& -\left[
    \tilde g_{\phi\phi} D^2(x,y) + 2\tilde g_{t\phi} D(x,y) N(x,y)\right.
  \nonumber\\
  & & \left.+\tilde g_{tt} N^2(x,y)\right]\ .
\end{eqnarray}
Equation (\ref{eqBernoulli}) is the Bernoulli equation we will now
consider. The solution $y(x)$ of the wind equations appears as the
level contour $E'$ of the surface $H(x,y)$. Notice that the classical
limits of $\beta'$, $\Theta'$ and $\omega'$ are not exactly the
corresponding $\beta$, $\Theta$ and $\omega$ parameters used by
\citet{sakurai:85}, who made the choice of normalizing
these quantities with the value of the gravitational potential at the
Alfv\'en point $GM/r_\mathrm{A}$ [see Eqs.~(11a), (11b) and (11c) of
\citet{sakurai:85}] whereas we used $c^2$ to make the definition of
these parameters more general.  However a simple relation applies
between Sakurai's and our parameters:
$\beta/\beta'=\Theta/\Theta'=\omega/\omega'=r_\mathrm{A}/r_g$ where
$r_g=GM/c^2$ is the gravitational radius.  This is also valid
for the classical limit of the Bernoulli function
$\tilde{H}(x,y)$ and the definition used by \citet{sakurai:85}.

Before studying $\tilde{H}(x,y)$ in the following section, we have to
note that this function is not defined everywhere in the region $x>0$,
$y>0$ as it is the case in the classical limit.  The function
$\mathcal{D}(x,y)$ must be strictly positive so that $\tilde{H}(x,y)$
is well defined and the velocities are not imaginary.  The domain
where this condition applies is determined in
appendix~\ref{appDomain}.  In the sub-Alfv\'enic region ($y>1$) this
domain lies always inside the light surface, its location is given
more precisely in appendix~\ref{appLC}.

\section{Description of the solutions}
\label{sec:descr}

\begin{figure}[htbp]
  \includegraphics[width=\hsize]{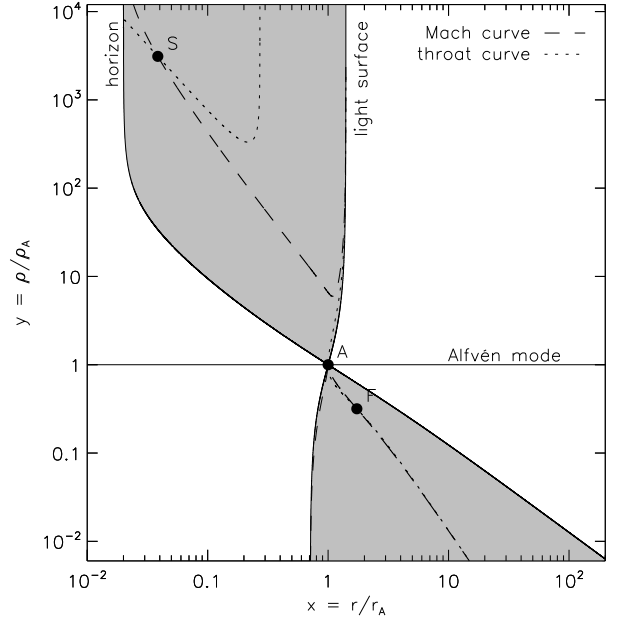}
  \caption{\textbf{Solution plane of the wind equations.}
    The gray region corresponds to the domain where the Bernoulli
    function is well defined. In the sub-Alfv\'enic region ($y>1$), it
    is limited by the light surface.  The thick line (Alfv\'en mode)
    separates the sub- and the super-Alfv\'enic modes.  The dashed
    line indicates the slow ($y>1$) and fast ($y<1$) mode Mach curves
    and the dotted line the gravitational throat curve.  The
    gravitational throat curve is very close to the fast mode Mach
    curve for this particular case.  The slow (S) and fast (F)
    critical points are the intersections of the Mach and throat
    curves. The Alfv\'en point is indicated by A.}
  \label{figDiagram1}
\end{figure}
\begin{figure}
  \includegraphics[width=\hsize]{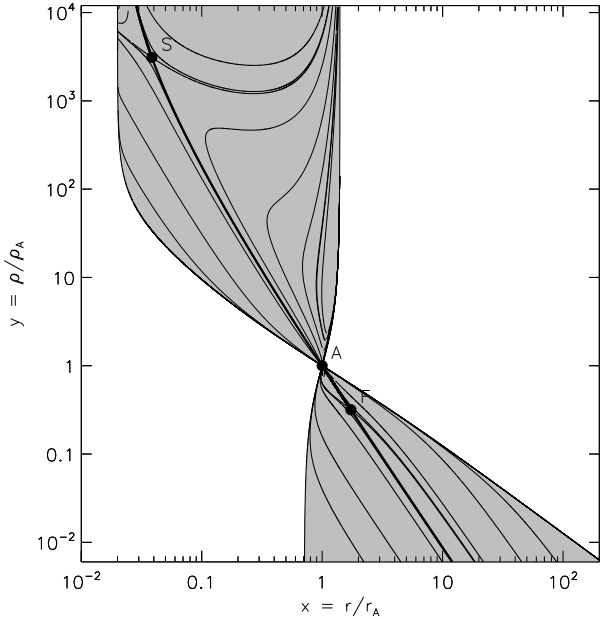}
  \caption{A few level contours of the Bernoulli
    function are shown. The physical solution (thick line) starts in
    the sub-Alfv\'enic region, crosses the slow mode Mach curve at the
    slow critical point (S) then reaches the Alfv\'en point (A) and
    enters the super-Alfv\'enic region where it crosses the fast mode
    Mach curve at the fast critical point (F). This calculation has
    been made for a Schwarzschild black hole with $m=0.01$ and the
    parameters ($\gamma=4/3$, $\Theta'=0.04$ and $\omega'=0.5$) have
    been chosen so that the different points are well separated.}
  \label{figDiagram2}
\end{figure}

\subsection{Properties of the Bernoulli function}

As described by \citet{sakurai:85} in the classical case, the
Bernoulli function $\tilde{H}(x,y)$ has the following properties:
\begin{enumerate}
\item At $y=1$ ($\rho=\rho_\mathrm{A}$) $\tilde{H}(x,y)$ diverges if
  $x \ne 1$ and remains finite if $x=1$ ($r=r_\mathrm{A}$). It means
  that all solutions going from the sub--Alfv\'enic region ($y>1$) to
  the super--Alfv\'enic region ($y<1$) must pass through the Alfv\'en
  point $x=y=1$ which is the only ``hole'' in the infinite ``wall''
  $y=1$.
\item Two important curves in the $x$--$y$ plane are the so called
  slow/fast mode Mach curve defined by
  \begin{equation}
    \frac{\partial\tilde{H}}{\partial y}(x,y) = 0
  \end{equation}
  (the slow mode corresponds to the sub-Alfv\'enic region $y>1$ and
  the fast mode to the super-Alfv\'enic region $y<1$) and the
  gravitational throat curve (so called by analogy with the de~Laval
  nozzle) defined by
  \begin{equation}
    \frac{\partial\tilde{H}}{\partial x}(x,y) = 0\ .
  \end{equation}
  At the intersections of these two curves, the function
  $\tilde{H}(x,y)$ is locally flat, corresponding to an X-type
  critical point (or O-type point).
\item All level contours of $\tilde{H}(x,y)$ going from $y\to +\infty$
  in the sub--Alfv\'enic region to $y\to 0^+$ in the super--Alfv\'enic
  region must cross these critical lines. They have to cross
  them simultaneously to be not interrupted. This means that the solution
  must pass through two critical points defined as the slow
  (respectively fast) critical point $x_\mathrm{s},y_\mathrm{s}$
  (resp.  $x_\mathrm{f},y_\mathrm{f}$), intersection of the slow
  (resp. fast) mode Mach curve and the gravitational throat curve.
  This imposes two new conditions for the solution:
  \begin{equation}
    \label{eqCondition}
    \tilde{H}\left(x_\mathrm{s},y_\mathrm{s}\right) 
    = E' 
    \qquad\mathrm{and}\qquad
    \tilde{H}\left(x_\mathrm{f},y_\mathrm{f}\right) 
    = E'\ .
  \end{equation}
\end{enumerate}
Figure~\ref{figDiagram1} shows the $x$--$y$ plane for a particular
choice of the parameters with the slow/fast mode Mach curve, the
gravitational throat curve.  Different level contours of the function
$\tilde{H}(x,y)$ are shown in Fig.~\ref{figDiagram2}.  The solution is
one of the level contours and passes through the slow and fast
critical points.  These figures are very similar to Fig.~1 of
\citet{sakurai:85} obtained in the classical case.  Notice that
\citet{begelman:94} restrict their study to cold flows.  A consequence
of this assumption is that a purely radial flow is a singular case
where the fast critical point is at infinite radius.  The fast point
moves inward to finite radii only if the flow diverges over-radially,
like in the situations we will study in the next section.  This is not
a necessary condition in our more general model.  Because we include
gravity and thermal pressure the fast point is always located at
finite distances even in a purely radial flow.

\subsection{Classification of wind solutions by dimensionless
  parameters}

\begin{figure}
  \includegraphics[width=\hsize]{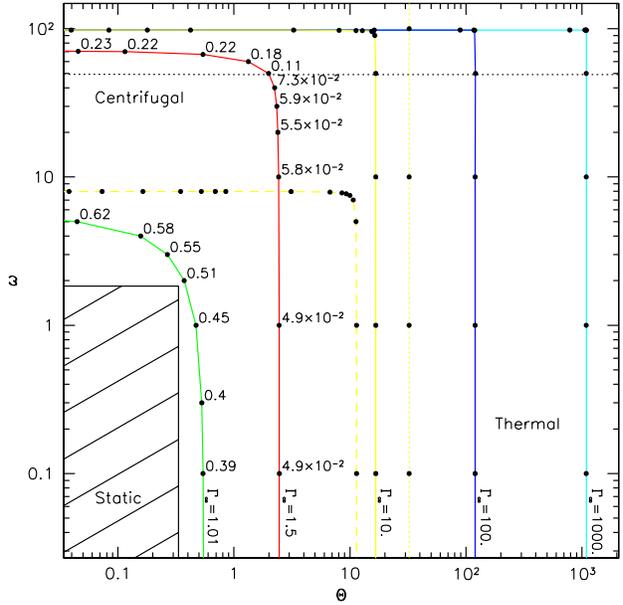}
  \caption{\textbf{Parameter space:} there are no wind solutions in the
    lower left part (``Static'': the limits are only approximatively
    indicated). The upper left part (``Centrifugal') corresponds to
    winds where the thermal pressure is negligible whereas in the
    lower right part (``Thermal'') the centrifugal force plays no
    essential role. For a Schwarzschild metric with $m=0.01$ (so
    $\omega'<0.98$ and $\omega<98$) we have computed several series of
    solutions with constant terminal Lorentz factor
    $\Gamma_{\infty}=1.01$, $1.5$, $10$, $100$ and $1000$ (solid
    lines). For $\Gamma_{\infty}=10$ we show also the case where
    $m=0.1$ (dashed line: $\omega<8$) and $m=0.001$ (dotted line:
    $\omega<998$). The dotted horizontal line corresponds to $\xi=1$
    (initially the electromagnetic and the matter energy fluxes are
    equal) for $m=0.01$. For $m=0.01$ and $\Gamma_{\infty}=1.01$ and
    $1.5$ we indicate at different positions the value of the
    efficiency $\mathit{eff}$ of the magnetic to kinetic energy
    conversion.}
  \label{figParameters}
\end{figure}

Equation~(\ref{eqBernoulli}) can formally be written as
\begin{equation}
  \tilde{H}\left(x,y\,; \beta',\Theta',\omega', m, g, \gamma\right) 
  = E'\ ,
\end{equation}
where $m$ stands for the parameters defining the metric (see
appendix~\ref{appMetric} for the definition of $m$ in usual cases) and
$g$ stands for the parameters fixing the geometry of the flux tubes
(i.e. defining $\tilde{s}(x)$).  We will for the moment restrict our
study to the case where $\tilde{s}=\mathrm{const.}=1$ and we fix the
adiabatic index to $\gamma=4/3$ for the rest of this paper. From the
two conditions at the slow/fast critical points (\ref{eqCondition}) we
can fix two parameters. In practice, following \citet{sakurai:85}, we
adjust $\beta'$ so that the value of $\tilde{H}$ is the same at the
two critical points and this value gives $E'$. Therefore all solutions
are determined by only three independent parameters: $\Theta'$,
$\omega'$ and $m$. Compared to the classical case studied by
\citet{sakurai:85}, there is one supplementary parameter: $m$. This is
due to the presence of a characteristic length scale related to the
structure of the space-time (typically the gravitational radius
$GM/c^2$) which has no classical counterpart.

The signification of these three parameters is clear: $m$ measures the
intensity of the gravitational field, $\Theta'$ gives the strength of
the thermal pressure (the pressureless case which is often considered
corresponds to $\Theta'=0$) and $\omega'$ measures the strength of the
centrifugal force in accelerating the wind, or equivalently the effect
of the magnetic field.

In the limit of the flat space-time (Minkowski metric) it is well
known that the slow point does not exist anymore and that the solution
cannot start at an arbitrary small radius because it cannot cross the
slow mode Mach curve without having an infinite derivative .  We
therefore cannot describe the region near the source but this is
clearly because gravity influences the solution significantly at small
radii and the Minkowski approximation breaks down.  In the case
without gravity, as in the classical case, a supplementary parameter
($\beta'$) must be fixed (for instance by fixing the mass flux).  In
many cases the physical conditions at the basis of the wind are a
complex and not well understood question, and some assumptions made
everywhere else in the flow are probably not valid here (like the
adiabaticity). In this context, just adopting a Minkowski metric and
including in the $\beta'$ parameter all the unknown physics fixing the
mass flux at the basis of the wind is more elegant than adopting a
Schwarzschild (or Kerr) metric and applying the solution up to the
source.

Before exploring the three--parameters space we have defined, it is
useful to express the relation between $m$, $\omega'$ and $\Theta'$
and more usual physical quantities.  If $m$ is the ratio of the
gravitational radius of the source over the Alfv\'en radius, we have
\begin{equation}
  m = 0.21\cdot
  \left(\frac{r_\mathrm{A}}{10^{6}\,\mathrm{cm}}\right)^{-1}
  \left(\frac{M}{1.4 M_\odot}\right)\ .
\end{equation}
The angular frequency $\Omega$ can be identified with the rotation
rate of the source and fixes the value of $\omega'$:
\begin{equation}
  \omega' = 0.11\cdot
  \left(\frac{\Omega}{10^4\,\mathrm{Hz}}\right)^2
  \left(\frac{r_\mathrm{A}}{10^6\,\mathrm{rm}}\right)^2\ .
\end{equation}
The local sound speed is defined by $(c^\mathrm{s}/c)^2=\gamma P/\rho
h$ so that $\Theta'$ can be related to the ratio of the sound speed at
the Alfv\'en point $c^\mathrm{s}_\mathrm{A}$ over the speed of light
(for $\gamma=4/3$):
\begin{equation}
  \Theta' = 1.0\cdot10^{-2}\cdot
  \left(\frac{c^\mathrm{s}_\mathrm{A}}{0.1 c}\right)^2
  \left(
    \frac{1-3\left(c^\mathrm{s}_\mathrm{A}/c\right)^2}{0.97}
  \right)^{-1}\ .
\end{equation}
The three parameters $\Theta'$, $\omega'$ and $m$ being fixed, the
solution of the wind equations is simply found by adjusting $\beta'$
so that the condition (\ref{eqCondition}) is fulfilled. For each step
with $\beta'$ fixed, the slow and fast critical points are determined
by a simple Newton-Raphson procedure. The exact expressions of
$\tilde{H}$ and its derivatives are used. We have explored in detail
the parameter space and the results are presented in
Fig.~\ref{figParameters}.  Notice that for a given $m$, $\omega'$ is
limited to the interval
\begin{equation}
  \omega'_\mathrm{min} \le \omega' \le \omega'_\mathrm{max}\ ,
\end{equation}
where $\omega'_\mathrm{max}$ is due to the condition that the Alfv\'en
point lies inside the light surface and $\omega'_\mathrm{min}$ is non
vanishing only in the case of the Kerr metric. In this metric there
are no solutions without rotation ($\omega'=0$) because the matter is
forced to rotate in the vicinity of the central source. The analytical
expressions of $\omega'_\mathrm{min}$ and $\omega'_\mathrm{max}$ are
given in appendix \ref{appMetric}. Here we have considered a
Schwarzschild metric with different values of $m$. We show the results
in $\Theta$--$\omega$ coordinates, where $\Theta=\Theta'/m$ and
$\omega=\omega'/m$ are the parameters used by \citet{sakurai:85}.
Like in the classical case, there are no wind solutions in the lower
left part of the plane (``Static'') because neither the centrifugal
force (magnetic acceleration) nor the thermal pressure are sufficient
to power the wind. The limits of this region are only approximatively
indicated. For a pure thermal wind ($\omega=0$) and a Schwarzschild
metric, the minimal value of $\Theta$ is given by
\begin{equation}
  \Theta_\mathrm{min} 
  = \left(\gamma-1\right)\frac{\frac{1}{\sqrt{1-2m}}-1}{m}\ ,
\end{equation}
tending towards $\gamma-1$ for $m\to0$, in agreement with the
classical case. In the pressureless case ($\Theta=0$) the minimal
value of $\omega$ tends towards $(3/2)^{3/2}$ for $m\to 0$, which also
corresponds to the limit given by \citet{sakurai:85}.  The upper left
part corresponds to winds where the centrifugal force dominates and
the lower right part corresponds to pure thermal winds.  In
Fig.~\ref{figParameters} we have plotted several solutions with
constant terminal Lorentz factor $\Gamma_{\infty}=1.01$, $1.5$, $10$,
$100$ and $1000$ for $m=0.01$ and in the particular case
$\Gamma_{\infty}=10$ we have also plotted the same curves for $m=0.1$
and $m=0.01$ to show the effect of varying the gravitational field.

\section{Efficiency of the magnetic to kinetic energy transfer}
\label{sec:eff}

\subsection{Expressions of the energy fluxes}

To study the efficiency of the winds computed in the previous section,
we need to express the different components of the energy flux along
the flow:
\begin{eqnarray}
  \dot{E}_\mathrm{matter} 
  & = & 
  \dot{m}\left(-h \frac{u_t}{c^2}\right)\nonumber\\
  & = &
  \dot{m}\left(
    c^2 + E + h \frac{\Omega}{c} \frac{u_\phi}{c}
  \right)\nonumber\\
  & = & 
  -\left(
    \tilde g_{tt} \tilde{u}^t + \tilde g_{t\phi}\frac{\tilde{u}^\phi}{c}
  \right)\ \tilde{h}(y)\ \dot{m} c^2\ ,\\
  \dot{E}_\mathrm{em}
  & = &
  \dot{m}\frac{\Phi^2\rho}{4\pi\dot{m}^2}
  \Omega\frac{g_{t\phi}^2-g_{tt}g_{\phi\phi}}{c^2}
  \left(\Omega u^t - u^\phi\right)\nonumber\\
  & = & 
  \dot{m}\left(
    \Omega L-h\frac{\Omega}{c}\frac{u_\phi}{c}
  \right)\nonumber\\
  & = &
  \sqrt{\omega'}x\tilde{\varpi}^2
  \frac{\tilde{h}(1)y}{M_\mathrm{A}^2}
  \left(
    \sqrt{\omega'}x\tilde{u}^t-\frac{\tilde{u}^{\rm \phi}}{c}
  \right)\ \dot{m}c^2\ ,\\
  \dot{E}_\mathrm{total}  
  & = & 
  \dot{E}_\mathrm{matter} + \dot{E}_\mathrm{em}\nonumber\\
  & = & 
  \dot{m}\left(c^2 + E + \Omega L\right)\nonumber\\
  & = & 
  -\left(
    \tilde g_{tt}(1) + \sqrt{\omega}'\tilde g_{t\phi}(1)
  \right)\ \frac{E'+1}{M_\mathrm{A}^2}\ \dot{m}c^2\ .
\end{eqnarray}
Notice that the matter part is made up of the rest-mass, kinetic and
internal energy of the matter. We define two parameters: the initial
baryonic load $\eta$
\begin{equation}
  \frac{1}{\eta} = \left.
    \frac{\dot{E}_\mathrm{matter}}{\dot{m}c^2}\right|_{x_0}
\end{equation}
and the ratio of the initial power injected in the electromagnetic
field over the initial power injected in the matter
\begin{equation}
  \xi = \left.
    \frac{\dot{E}_\mathrm{em}}{\dot{E}_\mathrm{matter}}
  \right|_{x_{0}}\ ,
\end{equation}
where $x_0$ is the radius where the wind starts. The value of $\xi$
depends only weakly on $x_0$, and we have arbitrary chosen $x_0=6m$
(or $r_0=6r_g$).  One sees that $\eta$ will be fixed by $\Theta'$ [via
the initial value of $\tilde{h}(y)$] whereas $\xi$ depends strongly on
$\omega'$.  Along the flow, the internal energy is converted into
kinetic energy which accelerates the wind. Therefore if there were no
magnetic field, the Lorentz factor at infinity would be
$\Gamma_\infty=1/\eta$. However, the magnetic field can also
contribute to the acceleration (when coupled with the rotation) and
depending on the efficiency of the conversion of electromagnetic into
kinetic energy, the terminal Lorentz factor can be larger than
$1/\eta$, with a maximum value (complete conversion) given by
\begin{equation}
  \Gamma_\infty^\mathrm{max} = \frac{1+\xi}{\eta}\ .
\end{equation}
In reality the conversion will never be complete and will be estimated
by the following fraction
\begin{equation}
  \mathit{eff} = 1
  - \frac{\left.\dot{E}_\mathrm{em}\right|_{\infty}}
  {\left.\dot{E}_\mathrm{em}\right|_{0}}
  = \frac{\eta\Gamma_{\infty}-1}{\xi}\ .
\end{equation}
\subsection{Inefficient conversion}
In the pressureless case ($\Theta'=0$) it is well known that the
magnetic to kinetic energy transfer is very inefficient for high
terminal Lorentz factors \citep{michel:69}
\begin{equation}
  \frac{\Gamma_\infty \dot{m} c^2}{\dot{E}_\mathrm{total}} 
  = \frac{1}{\Gamma_\infty^2}\ .
\end{equation}

The curves for constant terminal Lorentz factor in
Fig.~\ref{figParameters} show clearly that for highly relativistic
winds ($\Gamma_\infty=10$, $100$ and $1000$) the terminal Lorentz
factor $\Gamma_\infty$ is independent of $\omega$ (or equivalently of
$\xi$) which means that there can only be a tiny magnetic to kinetic
energy conversion.  When $\omega$ is very close to the maximal allowed
value and the outflow is Poynting flux dominated (corresponding to the
case where the Alfv\'en point is at the light surface radius) this
tendency is not valid anymore. In this region the terminal Lorentz
factor $\Gamma_\infty$ depends strongly on $\omega$ and is almost
independent of $\Theta$ (or equivalently of $\eta$).  However, even in
this case only a tiny fraction of the magnetic energy is converted
into kinetic energy.  The converted energy amount is great compared to
the initial energy in the matter part and therefore leads to a greater
increase in $\Gamma$ and $\dot E_\mathrm{matter}$ throughout the flow.
The efficiency $\mathit{eff}$ of the conversion is maximal in the
pressureless case ($\Theta=0$) but is still rather small.  In this
case the parameters $\eta$ and $\xi$ are given by $\eta \simeq 1$ and
$\xi \simeq \dot E_\mathrm{total}/\dot mc^2 - 1 =
\Gamma^{3}_{\infty}-1$ which corresponds to
\begin{equation}
  \mathit{eff} \simeq 
  \frac{1}{1+\Gamma_\infty+\Gamma_\infty^2}\ .
\end{equation}
These tendencies are still present in mildly relativistic winds, as
can be seen on Fig.~\ref{figParameters} for $\Gamma_\infty=1.5$ where
we have indicated the evolution of $\mathit{eff}$ along the curve. It
is only within the classical limit that the conversion becomes
important. This is shown in Fig.~\ref{figParameters} [The case
$\Gamma_\infty=1.01$ ($v_\infty=0.14\,c$)]. Here the efficiency of
conversion reaches $\sim~60\,\%$ in the pressureless case, which is in
agreement with the classical study of \citet{sakurai:85}.

\subsection{Efficient conversion}

All wind solutions showing (except in the classical limit) a very bad
efficiency of the electromagnetic to kinetic energy conversion have
been calculated for a particular geometry corresponding to
$\tilde{s}=\mathrm{const.}=1$ in our dimensionless units.  This
corresponds to magnetic flux tubes of constant opening angle.  Under
the assumption that the velocity is purely radial and constant at
infinity it is possible to predict analytically the asymptotic
behavior of the flow for any kind of geometry $\tilde{s}(x)$:
\begin{eqnarray}
  \tilde{u}^t & \to & \Gamma_\infty\ ,\nonumber\\
  \tilde{u}^r & \to & \sqrt{\Gamma_\infty^2-1}\ ,\nonumber\\
  \tilde{u}^\phi & \to & 0\ ,\nonumber\\
  y & \simeq &
  \frac{\sqrt{\beta'}}{\sqrt{\Gamma^2_{\infty}-1}}
  \frac{1}{\tilde{s}x^2}\ ,\nonumber
\end{eqnarray}
so that the asymptotic expressions of the energy fluxes are
\begin{eqnarray}
  \frac{\dot{E}_\mathrm{matter}}{\dot{m}c^2} & \to &
  \Gamma_\infty\ ,\\
  \frac{\dot{E}_\mathrm{em}}{\dot{m} c^2} & \to &
  \frac{\omega'\sqrt{\beta'}\tilde{h}(1)}{M_\mathrm{A}^2}
  \frac{1}{v_\infty}\frac{1}{\tilde{s}}\ . \label{eq:EemInf}
\end{eqnarray}
From the last equation, one sees that the magnetic to kinetic energy
conversion depends strongly on $\tilde{s}$. At infinity $\tilde{s}\to
0$ is unphysical because it would mean that the energy diverges. The
case where the opening angle is constant at infinity corresponds to
$\dot{E}_\mathrm{em}\to\mathrm{const.}> 0$ at infinity so that the
conversion is not complete and the case where the opening angle
diverges ($\tilde{s}\to +\infty$) gives $\dot{E}_\mathrm{em}\to 0$ so
that the conversion is complete. These results indicate that all
models considered in the previous section are inefficient due to a
particular choice of the geometry: $\tilde{s}=\mathrm{const}$. This
assumption is certainly correct at very large distance from the source
but the opening angle may have variations at smaller radii.
Equation~(\ref{eq:EemInf}) indicates that every region where the
opening angle increases is a region of efficient magnetic to kinetic
energy transfer.  This is in agreement with the results of
\citet{begelman:94}.

\begin{figure*}
  \includegraphics[bb=24 162 581 713,height=16.5cm,width=\hsize]{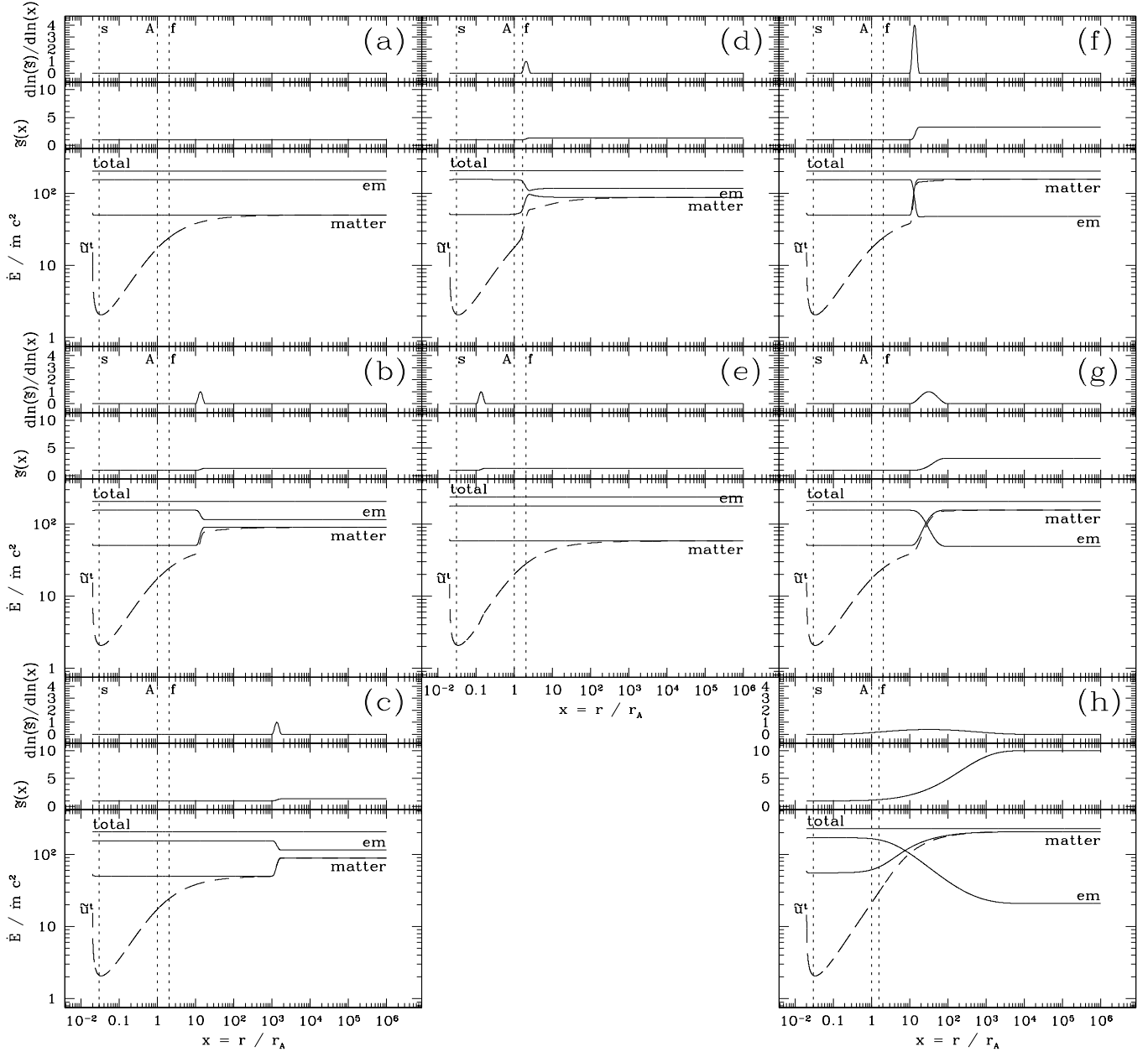}
  \caption{\textbf{Effect of the geometry on the efficiency of the
      magnetic to kinetic energy conversion:} we consider a
    Schwarzschild metric with $m=0.01$ and a wind solution
    characterized by $\Theta'=0.65$ and $\omega'=0.74$. This
    corresponds to an initial energy flux which is dominated by the
    electromagnetic part: $\xi=3.1$ for all models so that initially
    $75\ \%$ of the energy is magnetic. All solutions presented here
    have $\eta=2.\ 10^{-2}$ except for cases e ($\eta=1.7\ 10^{-2}$)
    and h ($\eta=1.8\ 10^{-2}$). The slow and fast critical points are
    located at $x_{s}=3.0\ 10^{-2}$ and $x_{f}=2.0$ except for cases d
    ($x_{f}=1.6$) and h ($x_{f}=1.6$).  On each figure
    $\mathrm{d}\log\tilde{s}/\mathrm{d}\log{x}$, $\tilde{s}(x)$ and
    the different components of the energy flux (matter/em and total)
    are presented as functions of the radius $x$. The ``Lorentz
    factor'' $\tilde{u}^\mathrm{t}$ is also shown (dotted line). Three
    vertical dotted lines show the location of the slow (s), the
    Alfv\'en (A) and the fast point (f). \textit{Case a:}
    $\tilde{s}=\mathrm{const.}=1$.  For this particular choice of the
    geometry, the conversion is extremely inefficient
    ($\mathit{eff}=3.8\ 10^{-4}$) and the terminal Lorentz factor
    equals $\Gamma_{\infty}=50=1/\eta$.  \textit{Case b:} $\tilde{s}$
    increases between $x_{1}=10$ and $x_{2}=18$ reaching a maximal
    slope $\mathrm{d}\log{\tilde{s}}/\mathrm{d}\log{x}=1$.  The
    efficiency improves a lot: $\mathit{eff}=0.26$ and
    $\Gamma_{\infty}=90$.  \textit{Case c:} same as b but $\tilde{s}$
    increases between $x_{1}=1000$ and $x_{2}=1800$ ($x_{2}/x_{1}$ is
    the same). It changes neither the efficiency nor the terminal
    Lorentz factor.  \textit{Case d:} same as b but $\tilde{s}$
    increases between $x_{1}=1.5$ and $x_{2}=2.7$ ($x_{2}/x_{1}$ is
    the same), i.e.  before the position of the fast point in the
    reference solution a.  Again the efficiency $\mathit{eff}=0.24$
    and the terminal Lorentz factor $\Gamma_{\infty}=88$ are almost
    unchanged.  Notice that the fast critical point has moved to be
    almost at $x_{1}$.  \textit{Case e:} same as b but $\tilde{s}$
    increases between $x_{1}=0.1$ and $x_{2}=0.18$ ($x_{2}/x_{1}$ is
    the same), i.e.  before the Alfv\'en point. The efficiency is
    again very low: $\mathit{eff}=1.2\ 10^{-4}$ and
    $\Gamma_{\infty}=58\simeq 1/\eta$.  \textit{Case f:} same as b but
    with a maximal slope of
    $\mathrm{d}\log{\tilde{s}}/\mathrm{d}\log{x}=4$. The efficiency is
    better: $\mathit{eff}=0.69$ and $\Gamma_{\infty}=157$.
    \textit{Case g:} same as b but the region where $\tilde{s}$
    increases is larger: $x_{1}=10$ and $x_{2}=100$. Again the
    efficiency is better: $\mathit{eff}=0.68$ and
    $\Gamma_{\infty}=156$. \textit{Case h:} we have considered a case
    where $\tilde{s}$ increases from $x_{1}=0.1$ to $x_{2}=10^{4}$
    with a maximal slope
    $\mathrm{d}\log{\tilde{s}}/\mathrm{d}\log{x}=0.4$. Almost $90\ \%$
    of the magnetic energy is converted into kinetic energy
    ($\mathit{eff}=0.88$) so that $\Gamma_{\infty}=206$.}
  \label{figAlpha}
\end{figure*}

To check that the geometry is really the key parameter governing the
efficiency of such winds we have computed some models using various
laws for the evolution of the opening angle $\tilde{s}(x)$. The
results are shown in Fig.~\ref{figAlpha} and confirm the previous
analysis. We have considered a Schwarzschild metric with $m=0.01$ and
a wind model characterized by $\omega'=0.97$ and $\Theta'=0.1$ (so the
energy flux is initially dominated by the electromagnetic energy
flux). We plot the different energy fluxes and the ``Lorentz
factor\footnote{For commodity, we use the expression ``Lorentz
  factor'' for $\tilde{u}^\mathrm{t}$ even if, strictly speaking, this
  should only be used at large distance of the source where the metric
  is very close to the Minkowski metric, i.e. for $x \gg 1$.  The
  correct expression of the Lorentz factor should be corrected with
  the lapse function, $\Gamma = \alpha \tilde{u}^\mathrm{t}$, where
  $\alpha=1/\sqrt{-\tilde{g}^{tt}} \to 1$ for $x \to +\infty$.}''
$\tilde{u}^\mathrm{t}$ as well as the geometrical function
$\tilde{s}(x)$ we have used in each case.  Figure~\ref{figAlpha}a
corresponds to the inefficient case $\tilde{s}=\mathrm{const.}=1$.
Figure~\ref{figAlpha}b corresponds to the case where $\tilde{s}$
increases in a region located between $x_1=10$ and $x_2=18$: the
magnetic to kinetic energy conversion is immediately better.  The
efficiency $\mathit{eff}$ increases also in geometries with different
shapes (Fig.~\ref{figAlpha}d,f,g,h) and different locations of the
$\tilde s>1$--region, provided that this region lies beyond the fast
point as shown by \citet{begelman:94}.  In this case
$\tilde{s}_{\infty}$ is the only relevant quantity which governs
$\mathit{eff}$.  A $\tilde s>1$--region within the fast point like in
Fig.~\ref{figAlpha}e does not increase $\mathit{eff}$ and is similar
to the purely radial case.

\begin{figure}
  \includegraphics[width=\hsize]{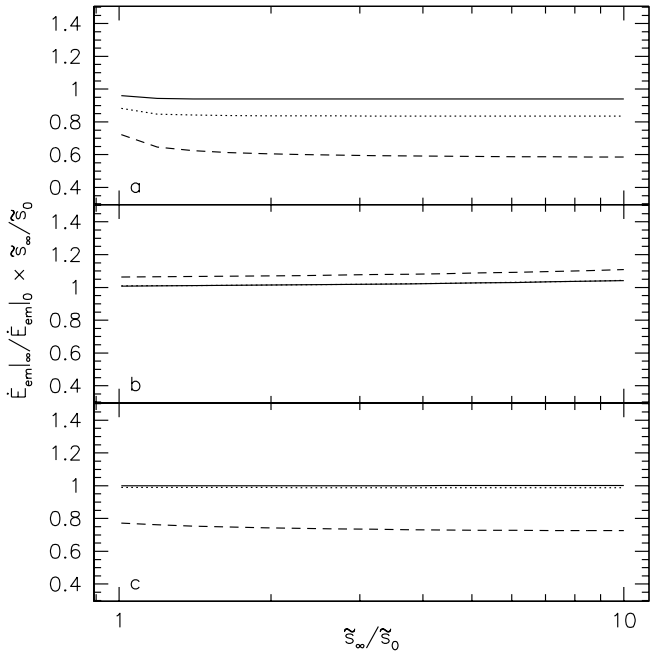}
  \caption{The influence of the geometry, gravitation and
    thermal energy upon the electromagnetic energy conversion.  The
    solid, dotted and dashed lines correspond to different
    gravitational field strengths, $m=0.001$, $m=0.01$ and $m=0.1$.
    \textbf{a)} The case for $\omega'=0.95\,\omega'_\mathrm{max}$,
    $\Theta'=0.01$ which is a cold Poynting-flux dominated outflow.
    \textbf{b)} Thermal energy dominated fireball:
    $\omega'=0.1\,\omega'_\mathrm{max}$, $\Theta'=10$.  The dotted
    line lies very close to the solid line and is not visible.
    \textbf{c)} Non-relativistic case where the rest mass dominates
    with $\omega'=0.1\,\omega'_\mathrm{max}$, $\Theta'=0.1$.}
  \label{fig:sstudy}
\end{figure}

\citet{begelman:94} showed that the electromagnetic energy flux
decreases like
\begin{equation}
\label{eq:Eem-s-rel}
  \frac{\left.\dot{E}_\mathrm{em}\right|_{\infty}}
    {\left.\dot{E}_\mathrm{em}\right|_{x_0}}
      = \frac{\tilde{s}(x_\mathrm{f})}{\tilde{s}_\infty}
\end{equation}
for a cold flow in Minkowski metric.  If the asymptotic regime is
already reached in the region where the opening angle increases,
Eq.~(\ref{eq:EemInf}) shows that this relation should still be valid in
the most general case, independent of the gravitational field or of
the initial amount of thermal energy.  To check the validity of this
result we consider 9 different $\omega',\theta',m$ combinations,
illustrating all possible situations and for each of them we compute
the evolution of the efficiency when varying $\tilde{s}_{\infty}$. As
Fig.~\ref{figAlpha} shows, the exact shape of the geometry is not
important, so we adopt a particular choice where $\tilde{s}$ rises
from $\tilde{s}_{0}=1$ to $\tilde{s}_{\infty}$ between $x_{1}=100$ and
$x_{2}=200$. This region lies always in the super-Alfv\'enic region,
which as discussed above is the condition for magnetic to kinetic
energy conversion.  Figure~\ref{fig:sstudy} shows the quantity
\begin{equation}
  \label{eq:ssq}
  \frac{\left.\dot E_\mathrm{em}\right|_{\infty}}
    {\left.\dot E_\mathrm{em}\right|_{x_0}}
  \cdot\frac{\tilde{s}_{\infty}}{\tilde{s}_{0}}
\end{equation}
plotted over $\tilde{s}_{\infty}/\tilde{s}_{0}$ for the 9 different
cases.  Notice that with our choice of geometry
$\tilde{s}(x_\mathrm{f})=\tilde{s}_{0}$.  One sees that gravity and
pressure changes the simple picture a bit.  In the cold cases
(Fig.~\ref{fig:sstudy}a and Fig.~\ref{fig:sstudy}c) the converted
energy fraction decreases for a stronger gravitational field and high
values of $m$.  On the other hand the gravitational field increases
the energy conversion by a small amount in the hot thermal dominated
case as seen in Fig.~\ref{fig:sstudy}b.  But (\ref{eq:Eem-s-rel})
remains still valid within a factor of 2.  For the cases of low
gravity and low thermal energy (solid lines in Fig.~\ref{fig:sstudy}a
and Fig.~\ref{fig:sstudy}c) the quantity (\ref{eq:ssq}) approaches $1$
as expected.

We can therefore conclude that the flow geometry \emph{always}
dominates the energy conversion and all other parameters play an only
minor role.

\section{Application to gamma--ray bursts}
\label{sec:grb}

Since the discovery of their optical afterglows gamma-ray bursts
(hereafter GRBs) have been known to be located at cosmological
distance. More than ten redshifts have been measured from $z=0.43$
(GRB\,990712) to $z=4.5$ (GRB\,000131).  The corresponding radiated
energy in the gamma-ray domain (20 -- 20000\,keV) ranges from
$5\cdot10^{51}\,$erg (GRB\,970228) to $2\cdot10^{54}\,$erg
(GRB\,990123) assuming isotropic emission.  Most sources that have
been proposed to explain such a huge release of energy in a few
seconds involve a rapidly rotating compact stellar-mass core. Among
them the two most popular are mergers of compact objects (neutron
stars binary or neutron star -- black hole systems) or collapses of
very massive stars to a black hole (collapsars)
\citep{meszaros:92,narayan:92,mochkovitch:93,woosley:93,paczynski:98}.
In both cases, the resulting system is a stellar mass black hole
surrounded by a thick torus made of stellar debris or of infalling
stellar material partially supported by centrifugal forces.  An other
interesting proposition \citep{usov:92,kluzniak:98,spruit:99}
associates GRBs with highly magnetized millisecond pulsars.  The
location of the detected optical counterparts, well inside their host
galaxy and possibly associated with star-forming regions, seems to
favor the collapsar scenario. However the other propositions cannot be
ruled out, at least for short bursts, for which no optical counterpart
has been detected yet.

Whatever the source is, the released energy must initially be injected
in a wind which eventually becomes relativistic. The existence of such
a relativistic wind has been directly inferred from the observations
of radio scintillation in GRB\,970508 \citep{frail:97} and is also
needed to avoid photon-photon annihilation. The absence of signature
of this last process in the BATSE spectra of GRBs implies very high
Lorentz factor for the wind: $\Gamma \sim 100$--$1000$
\citep{goodman:86,baring:95}. The second step consists in the
conversion of a fraction of the wind kinetic energy into gamma--rays,
probably via the formation of shocks within the wind itself
\citep{rees:94,daigne:98}. Such internal shocks are expected if the
wind is generated with a highly non uniform distribution of the
Lorentz factor so that rapid layers catch up with slower ones. In the
last step, the wind is decelerated when it interacts with the
environment of the source and the resulting external shock is
responsible for the afterglow observed in X-ray, optical and radio
bands.

The origin of the relativistic wind is the most complex of the three
steps in this scenario. Several proposals have been made but only few
calculations have been performed so that none appears to be fully
conclusive. However it is suspected that large magnetic fields play an
important role. In a previous paper \citep{spruit:01} we have
considered different possible geometries of magnetic fields in GRB
outflows and we have proposed that in many cases, dissipation of
magnetic energy by reconnection should occur. The model we have
presented in this paper allows us to investigate these questions in
more details. In particular we focus on the case where the outflow
generated by the central engine is initially Poynting flux dominated
(in the following, we assume that only $10\%$ of the energy is
initially injected in the matter). To be consistent with the
observations showing that at the beginning of the afterglow emission,
the matter flow is highly relativistic, we also impose that the
terminal Lorentz factor has a large value (in the following, we will
adopt $\Gamma_\infty=100$). This implies a reasonable efficiency of
the magnetic to kinetic energy conversion.  The goal of the study
presented in this section is to illustrate that there are geometries
allowing such an efficiency and to discuss the possibility of magnetic
reconnection in this scenario.

\begin{figure}
  \includegraphics[width=\hsize]{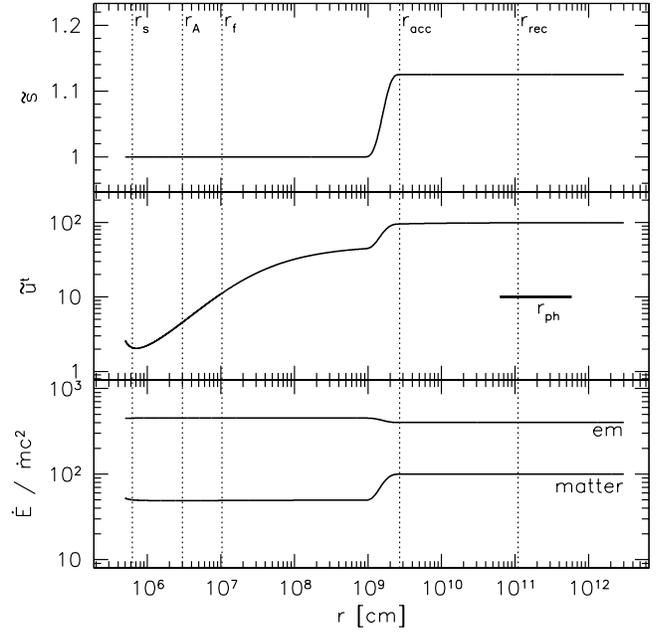}
  \caption{Geometry, ``Lorentz factor'' and energy fluxes for our
    example.  The vertical dotted lines mark the radii of the slow-,
    the Alfv\'en-, fast point and the acceleration radius.}
  \label{fig:rad_study}
\end{figure}

\citet{spruit:01} have shown that for typical GRB outflows the MHD
approximation is valid to very large distance ($\ga 10^{19}\,$cm)
which is the main assumption of our calculations. The second main
assumption -- the stationarity of the flow -- is of course less
justified in the case of GRBs. However we can estimate the time scale
to reach the stationary regime in our wind solutions as the time
needed by a particle starting from the basis of a flow line to reach
the Alfv\'en point:
\begin{equation}
  t_\mathrm{stat} = \frac{r_{a}}{c}\int_{x_{0}}^{1}
  \frac{\tilde{u}^{t}}{\tilde{u}^{r}}dx 
\end{equation}
(in the source frame). Let us estimate this time scale in a particular
case. We consider a Poynting flux dominated wind (we adopt $\xi=9.0$
so that only 10\,\% of the energy flux is initially injected in the
matter) with a moderately low initial baryonic load (we take
$\eta=1/50$). We impose that the terminal Lorentz factor is
$\Gamma_{\infty}=100$. If there were no magnetic to kinetic energy
conversion, the Lorentz factor at infinity would only be $1/\eta=50$.
In order to get a final Lorentz factor of $100$, we need to assume
that the geometry allows an efficiency $\mathit{eff} =
\left(\eta\Gamma_{\infty}-1\right)/\xi = 1/9$. We have shown in
Sect.~\ref{sec:eff} that this implies
$\tilde{s}_{\infty}/\tilde{s}_{0} \simeq \xi /
\left(1+\xi-\eta\Gamma_{\infty}\right)=1.125$.  For a given $m$, the
two other parameters $\Theta'$ and $\omega'$ are fixed by the values
of $\xi$ and $\eta$. We find that the following set of parameters:
$m=0.069$, $\Theta'=3.8$ and $\omega'=0.78$ fulfill the requirements
and corresponds to a reasonable value of the Alfv\'en radius
$r_\mathrm{A}$ and the angular frequency $\Omega$ in the case of a
millisecond pulsar-like source ($M=1.4M_\odot$) which is most likely
leading to an equatorial flow as we are considering here:
$r_\mathrm{A}=3.0\cdot10^{6}\,$cm and $\Omega = 8.8\cdot10^{3}\,$Hz.
Figure~\ref{fig:rad_study} shows the evolution of the ``Lorentz
factor'' and the electromagnetic and matter energy fluxes in this
case. We have assumed a simple geometry like those in
Sect.~\ref{sec:eff} where $\tilde{s}$ increases in a region located
between $x_{1}=300$ and $x_{2}=900$, well outside the fast critical
point radius. The corresponding time scale to reach the stationary
regime $t_\mathrm{stat}$ is between $\sim r_\mathrm{A}/c$ and $\sim 2
r_\mathrm{A} /c $, depending on the adopted value of the initial
radius $x_{0}$. As $r_\mathrm{A}/c = 10^{-4}\,$cm/s here, this is
compatible with the timescale of the variability observed in GRBs
profiles. This means that when the physical conditions at the basis of
the flow vary on a time scale $t_\mathrm{var} \ga 1\,$ms the flow
reacts instantaneously to reach a new stationary state corresponding
to the new boundary conditions.  Thus our calculation is a good
approximation for the relativistic wind of GRBs. If the wind produced
by the source lasts for a duration $t_\mathrm{w}$, our solution is
appropriate for the physical quantities within the corresponding shell
when it is located at radius $r$.

On the solution we present on Fig.~\ref{fig:rad_study}, the
acceleration occurs in two phases. First the initial thermal energy is
converted into kinetic energy, the magnetic energy remaining
unchanged.  This phase ends at $r\simeq 10^{9}\,$cm where
$\Gamma_{\infty}\simeq 1/\eta \simeq 50$. The second phase occurs in
the region where the opening angle increases. Here a magnetic to
kinetic energy conversion takes place.  We define the acceleration
radius $r_\mathrm{acc}$ as the radius where the flow reaches a Lorentz
factor of $\Gamma=0.95\Gamma_\infty$ and the acceleration can be
considered as finished.  The value of this radius is completely
dominated by the unknown flow geometry and equals $r_\mathrm{acc}
\simeq 2.7\cdot10^{9}\,$cm in this case.  Even if the location of the
region where the opening angle diverges would extend to higher radii
up to $10^{10}$--$10^{11}\,$cm, this radius is well below two other
important radii: the photosphere radius $r_\mathrm{ph}$ where the wind
becomes transparent and the reconnection radius $r_\mathrm{rec}$ where
the reconnection of the magnetic field should occur. These two radii
have been estimated in \citet{spruit:01}.  The photosphere radius is
the solution of
\begin{equation}
  1 = \int_{r}^{r+2\Gamma^2 c t_\mathrm{w}}
  \frac{\kappa \rho}{2\Gamma^2}dr
\end{equation}
and is independent of the duration of the burst $t_{w}$ as long as
$r_\mathrm{ph} \ll 2\Gamma^2 c t_\mathrm{w} =
6\cdot10^{14}\,\mathrm{cm}\cdot\left(\Gamma/100\right)^2
\left(t_\mathrm{w} / 1\,\mathrm{s}\right)$.  Here we have
$r_\mathrm{ph}=6.2\cdot10^{10}\,$cm for
$\dot{E}_\mathrm{total}=10^{51}\,$erg/s and
$r_\mathrm{ph}=5.9\cdot10^{11}\,$cm for
$\dot{E}_\mathrm{total}=10^{52}\,$erg/s.  This interval is marked by a
thick line on Fig.~\ref{fig:rad_study}.  As the remaining thermal
energy in the wind at such a large radius is very small, our adiabatic
wind solution applies up to the reconnection radius, where magnetic
dissipation starts.  This radius is given by
\begin{equation}
  r_\mathrm{rec} \simeq \frac{\pi c}{\epsilon \Omega}\Gamma^2
  \left(1+\frac{1}{\xi}\right)^{1/2}
\end{equation}
where $\epsilon<1$ is a numerical factor of order unity measuring the
reconnection speed in unit of the Alfv\'en speed. In our case we have
$r_\mathrm{rec}\simeq 1.1\cdot10^{11}\,\mathrm{cm}\cdot\epsilon$.  As
the magnetic energy flux is still 80\,\% of the total energy flux at
$r_\mathrm{rec}$, a very large amount of energy can possibly be
dissipated at this large distance.  Depending on the value of
$\dot{E}_\mathrm{total}$ and $\epsilon$, such reconnection events may
start when the wind is still optically thick (low $\epsilon$, high
$\dot{E}_\mathrm{total}$) or when the wind is already transparent
(high $\epsilon$, low $\dot{E}_\mathrm{total}$). As the dissipated
magnetic energy is probably first converted into thermal energy, the
consequences for the wind may be very different in these two cases.
(i) if the wind is optically thick, this injection of thermal energy
should be converted, at least partially (up to the photosphere radius)
into kinetic energy, leading to a third phase of acceleration; (ii) on
the other hand, if the wind is transparent, reconnection events could
directly contribute to the observed emission.  Notice that all the
radii we have computed are usually small compared to the typical
radius where internal shocks occur (with $\Gamma_{\infty}=100$)
\begin{equation}
  r_\mathrm{IS}\simeq 
  3\cdot10^{14}\,\mathrm{cm}\cdot
  \left(\frac{t_\mathrm{var}}{1\,\mathrm{s}}\right)\ ,
\end{equation}
where $t_\mathrm{var}$ is the typical time scale of the variability in
the initial distribution of the Lorentz factor and also small compared
to the deceleration radius where the external shock becomes efficient
(with $\Gamma_{\infty}=100$)
\begin{equation}
  r_\mathrm{dec}\simeq 
  5\cdot10^{16}\,\mathrm{cm}\cdot
  \left(\frac{E}{10^{52}\,\mathrm{erg}}\right)^{1/3}
  \left(\frac{n}{1\,\mathrm{cm}^{-3}}\right)^{-1/3}\ ,
\end{equation}
where $n$ is the density of the external medium and $E$ the total
energy of the wind at this radius. So these two ``standard''
mechanisms are not affected by the reconnection events. However the
relevant energy flux will be the kinetic energy flux at
$r_\mathrm{acc}$, possibly increased to a larger value if the
reconnection starts in the optically thick regime.

\section{Conclusions}
\label{sec:conclusions}

We have presented here a new formulation of the equations governing a
stationary axisymmetric MHD flow in the equatorial plane. This
formulation includes an exact treatment of all effects: thermal
pressure, gravity and arbitrary shape of the magnetic flux tubes. The
wind solution appears as the level contour of a Bernoulli-function
which passes through two particular points: the slow and fast critical
points.  It allows a direct comparison with the classical model of
\citet{weber:67}, in particular in the formulation given by
\citet{sakurai:85}. Thus the specifically relativistic effects are
easily identified.

We have used our model to extend the study of the magnetic to kinetic
energy conversion made by \citet{begelman:94}. We show that the main
parameter which fixes this efficiency is the shape of the magnetic
flux tubes. In the case of a constant opening angle, non-relativistic
flows have a good efficiency of the magnetic to kinetic energy
conversion but as soon as the terminal Lorentz factor is greater than
$\sim 1.5$, this efficiency decreases rapidly. Such relativistic winds
are not able to transfer a large fraction of their magnetic energy to
the matter. On the other hand, regions where the opening angle
diverges from the constant case are very efficient in converting
magnetic into kinetic energy, even in the ultra-relativistic case.
This is true as long as such regions are located beyond the fast
critical point.  Gravity and the thermal pressure play only a minor
role.

In Sect.~\ref{sec:grb}, we apply this model in the context of
gamma-ray bursts (GRBs hereafter).  In the case where the wind
produced by the source of GRBs is initially Poynting flux dominated,
we have shown that the efficiency of the acceleration strongly depends
on the geometry of the magnetic flux tubes. We found that a large
variety of situations is expected. If the magnetic tubes have the
possibility to diverge strongly from a constant opening angle, it is
possible that most of the energy is eventually in kinetic form. On the
other hand it is very likely that the magnetic to kinetic energy
conversion is incomplete and that the wind is still Poynting flux
dominated when it has reached its terminal Lorentz factor. We have
demonstrated on one example that such a wind can lead to very
promising situations compared to the standard picture: a large amount
of the magnetic energy can be dissipated at large radii by
reconnection. This reconnection can start when the wind is optically
thick or already transparent. So the large magnetic energy reservoir
could have two effects: a supplementary acceleration phase increasing
the final magnetic to kinetic energy conversion efficiency and/or a
direct contribution to the emission. These two possibilities will be
investigated in a future work.

\begin{acknowledgements}
  The authors would like to thank Dr. H.C.~Spruit for many stimulating
  discussions, important suggestions and reading the manuscript.
\end{acknowledgements}

\appendix

\section{Metric coefficients}
\label{appMetric}

The following table gives the metric coefficients in normalized units
in three cases: the Minkowski (M), Schwarzschild (S) and Kerr (K)
space-times.
\begin{equation}
  \begin{array}{|c|c|c|c|}
    \hline
    & \mathrm{M} & \mathrm{S} & \mathrm{K} \\
    \hline
    \hline
    \tilde g_{tt} & -1 & -1+\frac{2m}{x}  & -1+\frac{2m}{x} \\
    \tilde g_{t\phi} & 0 & 0  & -2a\frac{m^2}{x^2} \\
    \tilde g_{\phi\phi} & 1 & 1 &
    1+a^2\frac{m^2}{x^2}+2a^2\frac{m^3}{x^3}\\
    \tilde{g}_{rr} & 1 & \left(1-\frac{2m}{x}\right)^{-1} &
    \left(1-\frac{2m}{x}+a^2\frac{m^2}{x^2}\right)^{-1}\\
    \tilde{\varpi}^2 & 1 & 1-\frac{2m}{x} &
    1-\frac{2m}{x}+a^2\frac{m^2}{x^2} \\
    \sqrt{-\tilde{g}} & 1 & 1 & 1 \\
    \hline
    \hline
    x_\mathrm{h} & & 2 m & m\left(1+\sqrt{1-a^2}\right)\\
    x_\mathrm{e} & & & 2 m\\
    \hline
    \hline
    \sqrt{\omega'_\mathrm{min}} & 0 & 0 & 
    \frac{2 a m}{1+a^2 m^2+2 a^2 m^3}\\
    & & & \\
    \sqrt{\omega'_\mathrm{max}} & 1 & 1-2m & 
    \frac{2 a m^2 + \sqrt{1-2m+a^2m^2}}{1+a^2m^2+2 a^2 m^3}\\
    \hline
  \end{array}
\end{equation}
where $m=GM/r_{\rm a}c^2$ and $a=Jc/GM^2$ (where $J$ is the total
angular momentum of the black hole and $0\le a\le1$). The radii
$x_\mathrm{h}$ and $x_\mathrm{e}$ are respectively the radius of the
horizon and of the ergosphere. We consider only the case where $0\le m
\le \frac{1}{2}$ (the Alfv\'en point is outside the ergosphere). The
minimum value $\omega'_\mathrm{min}$ of $\omega'$ corresponds to the
condition $K_\mathrm{A} \ge 0$ (positive total angular momentum $L$)
and the maximum value $\omega'_\mathrm{max}$ corresponds to the
condition $M_\mathrm{A}^2>0$ (the Alfv\'en point must be inside the
light surface).

\section{The light surface}
\label{appLC}

The light surface is defined by (\ref{eqLC}) which limits the region
defined by
\begin{eqnarray}
  \xi^2(x) 
  &=& 
  a^2 m^2\omega' - 1 
  + 2\frac{m}{x}\left(1-a m\sqrt{\omega'}\right)^2
  + \omega' x^2
  \nonumber\\
  &\le& 0\ .
\end{eqnarray}
In the general case (with $\omega'\le\omega'_\mathrm{max}$, this
corresponds to a domain $x_{e} \le x^{-}_\mathrm{lc} \le x \le
x^{+}_\mathrm{lc}$ including the Alfv\'en point ($x_{a}=1$). In the
Minkowski case, $x^{-}_\mathrm{lc}=0$ and
$x^{+}_\mathrm{lc}=\frac{1}{\sqrt{\omega'}}$.

\section{Domain of definition of the Bernoulli function}
\label{appDomain}

\begin{figure*}
  \includegraphics[width=8.8cm]{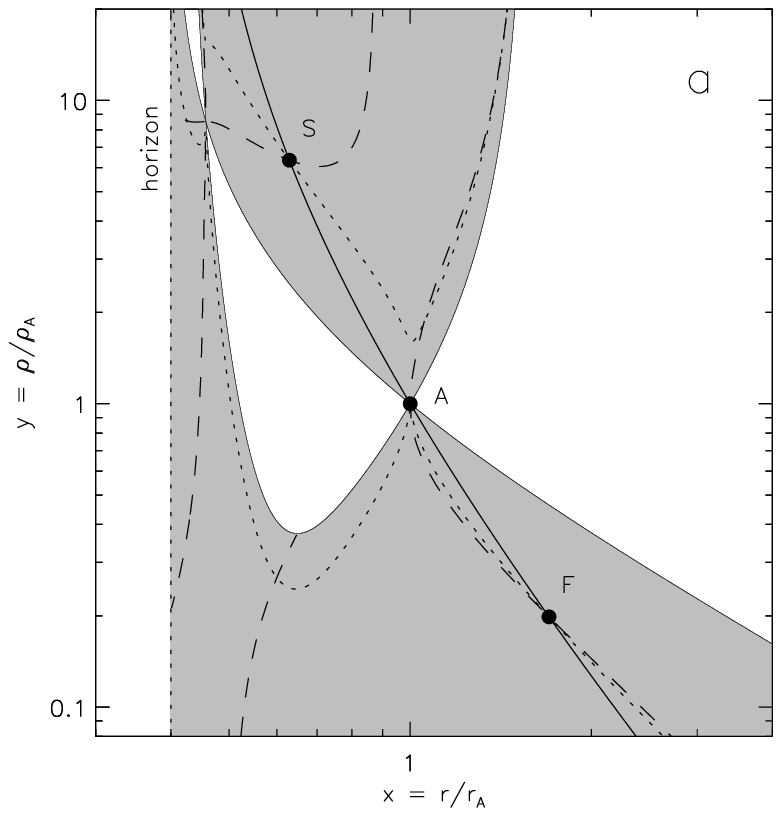} \hfill
  \includegraphics[width=8.8cm]{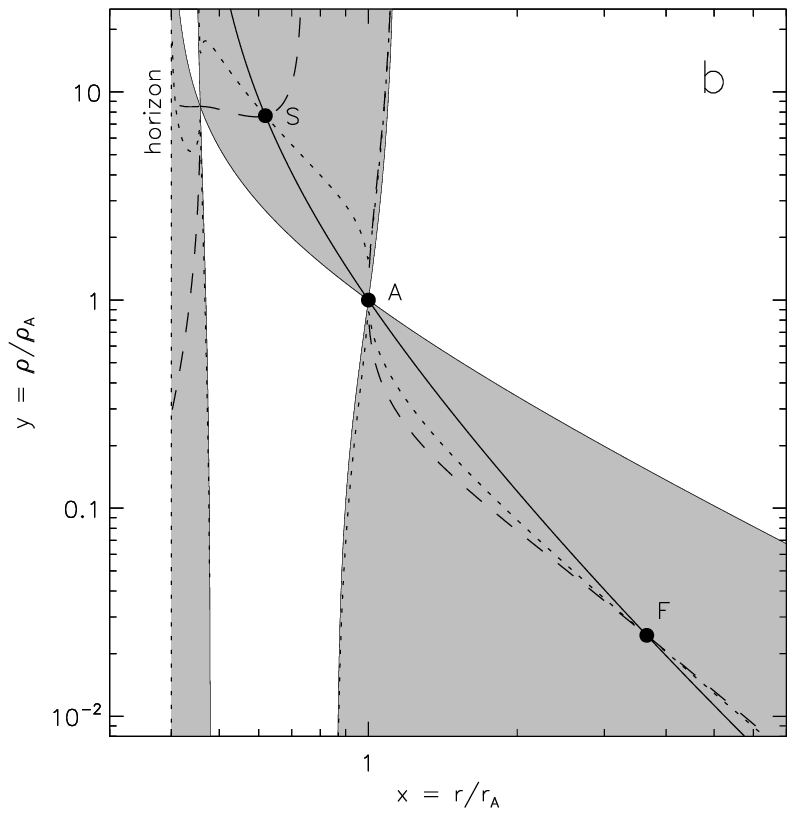}\\
  \raisebox{-8.8cm}{\includegraphics[width=8.8cm]{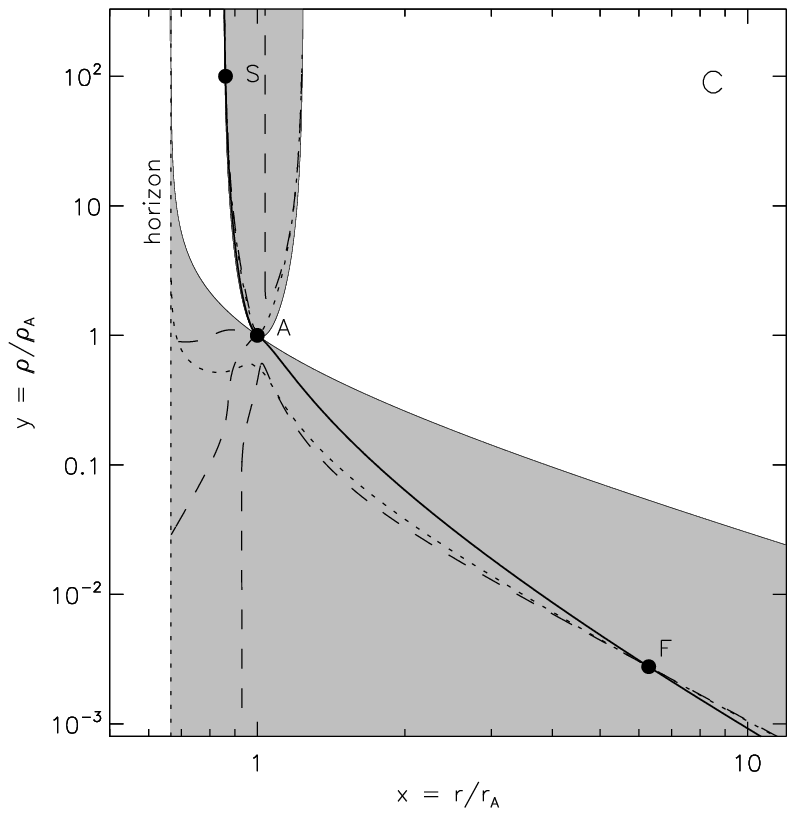}} \hfill
  \parbox[t]{8.8cm}{\caption{\textbf{Domain of definition of the
        Bernoulli function:} we consider a Schwarzschild black hole
      and draw this domain for different values of $m$ and $\omega'$
      (variations of $\Theta'$ correspond only to a contraction or a
      dilatation of the domain along the $y$ axis. We adopt here
      $\Theta'=1$).  \textit{Cases a and b: $m=0.2<m_{*}$.}  There are
      three possible configurations depending on the value of
      $\omega'$. Case a corresponds to $\omega'=0.3<\omega'_*$ and
      case b to $\omega'=0.5>\omega'_*$.  The case $\omega'=\omega'_*$
      looks very similar to case b and is not plotted.  We added the
      fast/slow mode Mach curves (dotted line) and the gravitational
      throat curve (dashed line) with the corresponding slow (S) and
      fast (F) critical point. The thick line is the solution passing
      through S,F and the Alfv\'en point A.  \textit{Case c:
        $m=m_{*}=1/3$.} This is the critical case where it is still
      possible to find a solution (here $\omega'=0.3$ and
      $\Theta'=1$).}}
  \label{figConfig}
\end{figure*}

The Bernoulli function $H\left(x,y\right)$ is defined for ${\cal
  D}(x,y)>0$ which gives the following condition:
\begin{equation}
  \label{eqDomain}
  A(x) Y^2 - 2 B(x) Y + C(x) > 0
\end{equation}
where $Y=\tilde{h}(1) y / \tilde{h}(y)$ is a function of $y$ and
$\Theta'$ only and $A$, $B$ and $C$ are functions of $x$, $\omega'$,
$a$ and $m$ only (notice that it is completely independent of the
function $\tilde{s}(x)$). The function $Y(y)$ is strictly increasing
from $Y(0)=0$ to $Y(+\infty)=+\infty$ with $Y(1)=1$. So we can focus
to the pressureless case $\Theta'=0$ and $Y=y$, all other cases
$\Theta'>0$ corresponding only to a contraction of the domain along
the $y$--axis. The coefficients $A$, $B$ and $C$ are given by
\begin{eqnarray}
  A(x) & = & -\left(\tilde{\varpi}^2x\right)^2\xi^2(x)\\
  B(x) & = & M_\mathrm{A}^2\left(\tilde{\varpi}^2 x\right)^2\\
  C(x) & = & \tilde{\varpi}^2\left\lbrace \tilde{\varpi}^2(1)
    \xi^2(x)\right.\\\nonumber
  & & \left. - M_\mathrm{A}^2 \left(
      \tilde g_{tt}(1)\tilde g_{\phi\phi} x^2 
      + 2\tilde g_{t\phi}(1)\tilde g_{t\phi} x 
      + \tilde g_{\phi\phi}(1)\tilde g_{tt}
    \right)\right\rbrace
\end{eqnarray}
The radius $x$ being fixed, the equation (\ref{eqDomain}) has 1 or 2
positive roots delimiting the domain where $H(x,y)$ is well defined.
Many configurations are possible for a Kerr metric and we do not
specify them in details here. We discuss only the case of the
Schwarzschild metric. Four different configurations are possible
depending on $m$ (with a critical case at $m_*=1/3$) and $\omega'$
(with a critical case at $\omega'=27 m^2 \left(1-2m\right)^2$). This
is illustrated on Fig.~\ref{figConfig}.

\bibliographystyle{aa}
\bibliography{h2857}

\begin{thebibliography}{26}
\expandafter\ifx\csname natexlab\endcsname\relax\def\natexlab#1{#1}\fi

\bibitem[{Baring(1995)}]{baring:95}
Baring, M. 1995, Ap\&SS, 231, 169

\bibitem[{Begelman \& Li(1994)}]{begelman:94}
Begelman, M.~C. \& Li, Z.-Y. 1994, ApJ, 426, 269

\bibitem[{Bekenstein \& Oron(1978)}]{bekenstein:78}
Bekenstein, J.~D. \& Oron, E. 1978, Physical Review D, 18, 1809

\bibitem[{Camenzind(1986{\natexlab{a}})}]{camenzind:86a}
Camenzind, M. 1986{\natexlab{a}}, A\&A, 156, 137

\bibitem[{Camenzind(1986{\natexlab{b}})}]{camenzind:86b}
---. 1986{\natexlab{b}}, A\&A, 162, 32

\bibitem[{Camenzind(1987)}]{camenzind:87}
---. 1987, A\&A, 184, 341

\bibitem[{Daigne \& Mochkovitch(1998)}]{daigne:98}
Daigne, F. \& Mochkovitch, R. 1998, MNRAS, 296, 275

\bibitem[{Frail {et~al.}(1997)Frail, Kulkarni, Nicastro, Feroci, \&
  Taylor}]{frail:97}
Frail, D., Kulkarni, S., Nicastro, S., Feroci, M., \& Taylor, G. 1997, Nature,
  389, 261

\bibitem[{Goldreich \& Julian(1970)}]{goldreich:70}
Goldreich, P. \& Julian, W. 1970, ApJ, 160, 971

\bibitem[{Goodman(1986)}]{goodman:86}
Goodman, J. 1986, ApJL, 308, L47

\bibitem[{Kennel {et~al.}(1983)Kennel, Fujimura, \& Okamoto}]{kennel:83}
Kennel, C.~F., Fujimura, F.~S., \& Okamoto, I. 1983, J. Astrophys. Geophys.
  Fluid Dyn., 26, 147

\bibitem[{Kluzniak \& Ruderman(1998)}]{kluzniak:98}
Kluzniak, W. \& Ruderman, M. 1998, ApJL, 505, L113

\bibitem[{M\'esz\'aros \& Rees(1992)}]{meszaros:92}
M\'esz\'aros, P. \& Rees, M. 1992, MNRAS, 257, 29

\bibitem[{Michel(1969)}]{michel:69}
Michel, F.~C. 1969, ApJ, 158, 727

\bibitem[{Mochkovitch {et~al.}(1993)Mochkovitch, Hernanz, Isern, \&
  Martin}]{mochkovitch:93}
Mochkovitch, R., Hernanz, M., Isern, J., \& Martin, X. 1993, Nature, 361, 236

\bibitem[{Narayan {et~al.}(1992)Narayan, Paczy\'nski, \& Piran}]{narayan:92}
Narayan, R., Paczy\'nski, B., \& Piran, T. 1992, ApJL, 395, L83

\bibitem[{Okamoto(1978)}]{okamoto:78}
Okamoto, I. 1978, MNRAS, 185, 69

\bibitem[{Paczy\'nski(1998)}]{paczynski:98}
Paczy\'nski, B. 1998, ApJL, 494, L45

\bibitem[{Parker(1958)}]{parker:58}
Parker, E.~N. 1958, ApJ, 128, 664

\bibitem[{Rees \& M\'esz\'aros(1994)}]{rees:94}
Rees, M. \& M\'esz\'aros, P. 1994, ApJL, 430, L93

\bibitem[{Sakurai(1985)}]{sakurai:85}
Sakurai, T. 1985, A\&A, 152, 121

\bibitem[{Spruit(1999)}]{spruit:99}
Spruit, H.~C. 1999, A\&A, 341, L1

\bibitem[{Spruit {et~al.}(2001)Spruit, Daigne, \& Drenkhahn}]{spruit:01}
Spruit, H.~C., Daigne, F., \& Drenkhahn, G. 2001, A\&A, 369, 694

\bibitem[{Usov(1992)}]{usov:92}
Usov, V.~V. 1992, Nature, 357, 472

\bibitem[{Weber \& Davis(1967)}]{weber:67}
Weber, E.~J. \& Davis, L.~J. 1967, ApJ, 148, 217

\bibitem[{Woosley(1993)}]{woosley:93}
Woosley, S.~E. 1993, ApJ, 405, 273

\end{thebibliography}

\end{document}